\documentclass [12pt,preprint]{aastex} 
\usepackage{epsfig} %
\begin{document}
 \voffset-1cm 
\newcommand{\gsim}{\hbox{\rlap{$^>$}$_\sim$}} 
\newcommand{\lsim}{\hbox{\rlap{$^<$}$_\sim$}}

\title{On re-brightening afterglows of XRFs, Soft GRBs and GRB081028} 

\author{Shlomo Dado\altaffilmark{1} and Arnon Dar\altaffilmark{2}}

\altaffiltext{1}{dado@phep3.technion.ac.il\\
Physics Department, Technion, Haifa 32000,
Israel}
\altaffiltext{2}{arnon@physics.technion.ac.il\\
Physics Department, Technion, Haifa 32000, Israel}

\begin{abstract}

It has been claimed recently that Swift has captured for the first time a 
late-time afterglow re-brightening of clear nonflaring origin after the 
steep decay of the prompt emission in a long gamma-ray burst (GRB), which 
may have been produced by a narrow jet viewed far off-axis. However, this 
interpretation of the observed re-brightening of the X-ray afterglow (AG) 
of GRB081028 is unlikely in view of its large equivalent isotropic 
gamma-ray energy. Moreover, we show that the late-time re-brightening of 
the AG of GRB081028 is well explained by the cannonball (CB) model of GRBs 
as a synchrotron flare emitted when the jet that produced the GRB in a 
star formation region (SFR) in the host galaxy crossed the SFR boundary 
into the interstellar medium or the halo of the host galaxy. We also show 
that all the other observed properties of GRB081028 and its afterglow are 
well reproduced by the CB model. On the other hand, we demonstrate that 
far-off axis GRBs, which in the CB model are `soft' GRBs and XRFs and 
consequently have much smaller isotropic equivalent gamma-ray energies, 
have slowly rising afterglows with a late-time power-law decay identical 
to that of ordinary GRBs, in good agreement with the CB model predictions.

\end{abstract} 


\section{Introduction}

A smoothly rising X-ray afterglow (AG) following the steep decay of the 
prompt emission in a long gamma ray burst (GRB) was first observed in 
GRB970508 with the Narrow Field Instrument (NFI) aboard the BeppoSAX 
satellite (Piro et al.~1998). Contemporaneous optical observations with 
ground based telescopes detected a corresponding re-brightening of its 
optical afterglow beginning around $6\times 10^4$ sec after burst (e.g., 
Galama et al.~1998 and references therein). This re-brightening was 
interpreted by Piro et al.~(1998) as due to a delayed activity of the 
central GRB engine.  Chiang \& Dermer~(1997) and Dar \& De R\'ujula~(2000) 
showed that a rise in the afterglow of a GRB produced by a jet of highly 
relativistic plasmoids (Shaviv \& Dar 1995) can result also from its 
deceleration if its viewing angle was initially far outside its 
relativistic beaming cone (far off-axis viewing angle). Both 
interpretations, however, failed to reproduce well enough the lightcurve 
of the AG of GRB970508. Consequently, Dado, Dar \& De R\'ujula (hereafter 
DDD) 2002 have investigated alternative explanations and showed that the 
late-time re-brightening of the AG of GRB970508 could be explained by the 
encounter of the jet with a density bump such as that expected at the 
boundary of the star formation region (SFR) where the GRB took place. A 
similar re-brightening of the X-ray afterglow of GRBs was later observed 
with the Swift X-ray telescope (XRT) in a few GRBs, e.g., in GRB060614 
(Mangano et al.~2007) and in several other GRBs where gaps in the data 
sampling and low statistics prevented a firm conclusion (e.g., GRBs 
051016B and 070306, Swift lightcurve repository, Evans et 
al.~2007,2009).

Overlooking the above, Margutti et al.~(2009) have recently claimed that 
Swift has captured for the first time a smoothly rising X-ray 
re-brightening of clear non-flaring origin after the steep decay of the 
prompt emission in a long gamma-ray burst GRB081028, which may have been 
produced by a narrow jet viewed far off-axis. But, the `far off-axis 
viewing' of GRB081028 is not supported by its large equivalent isotropic 
gamma-ray energy: The radiation emitted by a highly relativistic jet with 
a bulk motion Lorentz factor $\gamma\!\gg\!1$, which is observed from a 
small angle ($\theta\!\ll\!1$) relative to its motion, is amplified by a 
factor $\delta^3$ resulting from relativistic beaming and Doppler 
boosting, where the Doppler factor $\delta$ is well approximated by 
$\delta(t)\!=\![\gamma(t)\,(1-\beta(t)\, cos\theta]^{-1}\!\approx 
2\!\gamma/(1\!+\!\gamma^2\,\theta^2)$ (e.g. Shaviv \& Dar 1995; Dar 1998). 
Hence, for the typical viewing angle of GRBs, $\theta\!\sim\!1/\gamma$, 
$\delta\!\sim\! \gamma$, and the observed GRB fluence is amplified by a 
factor $\gamma^3$ relative to that in the jet rest frame. For `far 
off-axis viewing' angles where $\gamma^2\,\theta^2\!\gg\!1$, this 
amplification of $E_{iso}$ is reduced by a factor 
$[(1\!+\!\gamma^2\,\theta^2)/2]^{-3}$. For example, for $\theta\!=\! 
3/\gamma$, $E_{iso}$ is reduced by a factor $8\times 10^{-3}$ and GRBs 
viewed at $\theta\geq 3/\gamma$ appear as X-ray flashes (XRFs) rather than 
ordinary GRBs (see, e.g., Dar \& De R\'ujula 2000; DDD2004). From the 
observations of GRB081028, with the Burst Alert Telescope (BAT) aboard 
Swift, Margutti et al.~(2009) inferred that $E_{iso}\!=\!(1.1\!\pm\!0.1) 
\times 10^{53}$ erg, which is typical of ordinary GRBs and is inconsistent 
with the far off-axis interpretation of the re-brightening of its AG.
 
Despite the many serious discrepancies between the observed properties of 
GRB081028 and the predictions of the fireball (FB) model, which were 
pointed out by Margutti et al.~(2009) and are quite common in many other 
GRBs, alternative models of GRBs were ignored in their paper. One such 
model is the cannonball (CB) model, which has been shown repeatedly to 
reproduce well the main observed properties of GRBs and their AGs 
(DDD2009a and references therein). Thus, in this paper, we demonstrate 
that the main observed properties of GRB081028 and its afterglow are well 
reproduced by the CB model of GRBs. In particular, we show that the 
observed late-time re-brightening of the AG of GRB081028 is well explained 
as being a late-time synchrotron radiation (SR) flare. Such flares are 
expected to be emitted by the jet which produced the GRB, when it crosses 
the boundary of the star formation region, where presumably the GRB took 
place, into the ISM or the halo of the host galaxy. Moreover, we show that 
GRBs which have much larger viewing angles than those of ordinary GRBs, 
and thus appear as X-ray flashes (XRFs) or `soft' GRBs, have rather slowly 
rising late-time afterglows which are well described by the CB model.

\section{ICS and SR flares in the CB model}

\subsection{ICS flares}
The many similarities between the prompt emission pulses in gamma ray 
bursts (GRBs) and X-ray flares during the fast decay phase of the prompt 
emission and the early afterglow suggest a common origin. In the CB model 
of long GRBs this common origin is bipolar ejection of highly relativistic 
plasmoids (CBs) following mass accretion episodes of fall-back matter on 
the newly born neutron star or black hole in core-collapse supernova (SN) 
explosions akin to SN 1998bw (e.g., DDD2002; Dar \& De R\'ujula~ 2004; 
DDD2009a and references therein).  Both, the prompt pulses and early-time 
X-ray flares are produced by the thermal electrons in the CBs by inverse 
Compton scattering (ICS) of photons emitted/scattered into a cavity 
created by the wind/ejecta blown from the progenitor star long before the 
GRB. The prompt keV-MeV pulses in long GRBs are produced by CBs ejected in 
the early episodes of mass accretion. As the accretion material is 
consumed, one may expect the engine's activity to weaken.  X-ray flares 
during the decay of the prompt emission and the early afterglow phase are 
produced in such delayed episodes of mass accretion, which result in 
ejections of CBs with smaller Lorentz factors.

In the CB model, the lightcurve of ICS pulses/flares 
is given approximately by (DDD2009a):
\begin{equation}
E\, {d^2N_\gamma\over dt\,dE}(E,t)\approx
A\, {t^2/\Delta t^2  \over(1+t^2/\Delta t^2)^2}\,
E^{-\beta_g}\, e^{-E/ E_p(t)}\,,
\label{ICSlc}
\end{equation}
where $t\!=\!T\!-\!T_i$ 
with $T$ being the time after trigger and $T_i$ the beginning time of 
the pulse/flare
after trigger.  $A$ is a constant which depends on the CB's baryon number,
Lorentz and Doppler factors, on the density
of the glory light and on the redshift and distance of the GRB.
For $\beta_g$=0 (thin thermal spectrum of the glory photons),
$E_p(t)$ is the peak energy of $E\, d^2N_\gamma/ dE\, dt$ at time $t$.
Thus, in the CB model, each ICS pulse/flare in the GRB lightcurve
is described by four parameters, $A,$
$\Delta t(E),$  $E_p(t)$ and the
beginning time of the pulse $T_i$ when $t$ is taken to be 0.

The temporal behaviour of $E_p(t)$ depends 
on the self-absorption properties of the CBs and the spatial distribution 
of the seed photons in the glory which are not well known. 
Roughly, the peak energy $E_p(t)$ is given by (DDD2009a):
\begin{equation}
E_p(t)\approx  E_p(0)\, {t_p^k \over t^k+t_p^k}\,,
\label{PeakE}
\end{equation}
with $t_p$ being the time (after the beginning of the flare) when the ICS
photon count rate reaches its peak value and $2 \lsim k\lsim 3$
in order to accommodate the observed time dependence of $E_p$
at late time in several single-pulse GRBs.
For $E\!\ll\!E_p$, 
$E_p(t_p)\!=\!E_p$, where $E_p$ is the
peak energy of the time-integrated spectrum of the ICS 
pulse/flare (DDD2009a).

If absorption in the CB is dominated by free-free transitions then 
roughly, 
$\Delta t(E)\!\propto\! E^{-0.5}$, and for $E\!\ll\!E_p$ the lightcurve
of an ICF is approximately a function of $E\,t^2$ (the `$Et^2$
law'), with a peak at $t\!=\!\Delta t$, a full width at half maximum, FWHM
$\!\approx\!2\,\Delta t$ and a rise time from half peak value to peak
value, ${\rm RT\!\approx\! 0.30\,FWHM}$ independent of E.  Note that the
approximate $E\,t^2$ law makes the late decline sensitive only to the
product $E_p\, t_p^2$ and not to their individual values.

The evolution of the effective photon spectral index 
$\Gamma(t)=1+\beta$ of {\it isolated}
ICS pulses/flares is given approximately (DDD2008) by, 
\begin{equation} 
\Gamma(t)\approx 1+ {d\, log(F_\nu)\over d\,(logE )}\approx 
1+\beta_g+{E\,(t^2+t_p^2)\over E_p(0)\,t_p^2} \sim a+b\, t^2\, . 
\label{GammaICF} 
\end{equation} 
The $\!\sim\!t^2$ rise of $\Gamma$ during 
the fast decline phase of an ICS pulse/flare stops when a second 
pulse/flare or the power-law tail of SR flare (DDD2009a) that follows 
the ICS pulse/flare begins to dominate $F_\nu$ (DDD2008). Because of the 
exponential decay of the ICS pulses/flares, when the SR emission takes over 
the photon spectral index drops sharply to its constant typical SR value 
$\Gamma_X\!\sim\! 2.1$ (DDD2002).

\subsection{SR flares}

Synchrotron radiation begins when the jet of CBs 
crosses  the wind/ejecta which was blown from the progenitor star long 
before the GRB and continues during its deceleration in the interstellar 
medium. During these phases the electrons of the ionized gas in front of 
the CBs, which are swept in and Fermi-accelerated by the CBs turbulent 
magnetic fields, emit synchrotron radiation (SR) that dominates the 
prompt optical flares (DDD2009a; Dado \& Dar~2009a,b) and the 
following broad band afterglow emission. ICS of the SR radiation by the 
same electrons dominates the `prompt' high energy (GeV) emission that takes 
place simultaneously with the `prompt' optical emission (Dado \& Dar 
2009b). Assuming a canonical density profile of the wind/ejecta,
$n(r)\!\propto\! \Theta(r\!-\!r_w)\,
e^{\!-\!2l/(r\!-\!r_w)}/(r\!-\!r_w)^2\, $, 
where $r_w$ is the internal radius of the wind/ejecta, $r\!-\!r_w\!=\!l$ 
at max $n$ 
and $\Theta$ is a step function equal to zero for negative argument
and to 1 for positive argument, the early time SR flares (SRFs) have the 
lightcurve shape (DDD2009a, Dado \& Dar~2009a):
\begin{equation}
F_\nu[t] \propto  {e^{-a_w/t}\,
t^{1-\beta} \over t^2+t_{exp}^2}\, \nu^{-\beta}\!\rightarrow
        \! t^{\!-\!(1\!+\!\beta)}\, \nu^{\!-\!\beta}\, ,
\label{SRF}
\end{equation}
where $t\!=\!T\!-\!T_w$, $T$ is the observer time
after trigger and $T_w$ is the observer time when the CB reach 
the wind/ejecta at $r_w$. Such a lightcurve has a flare like shape, 
i.e., a fast rise followed by a power-law decay. Since the X-ray band  
is well above the `bend' frequency (DDD2009a), the X-ray flare has 
a temporal decay index 
$1\!+\!\beta_X\!\approx\!1\!+\!p/2\!\sim\! 2.1$ where $p$ is the injection 
spectral 
index
of the Fermi accelerated electrons (DDD2002).
The beginning of the `prompt' 
keV-MeV synchrotron emission that follows the ICS pulse/flare  is usually 
hidden 
under it. But, because of the exponential decay of the ICS pulse/flare,
the tail of the SR flare, which  
decays like $F_\nu \! \propto\!
t^{\!-\!\Gamma}\,\nu^{\!-\!\Gamma\!+\!1)}\,,$  takes over and
becomes visible in many GRBs (DDD2009a).

Long GRBs are mostly produced in core collapse supernova
explosions (DDD2009a and references therein) which usually take  
place in star formation regions (SFRs). In that case 
Eq.~(\ref{SRF})  describes also the late-time SRF which is produced
when the jet crosses the boundary of the SFR into the ISM or 
the halo of the host galaxy. 
This usually takes place around $t\!\approx\! 
(1+z)\,R_{SFR}/2\,c\,\gamma^2$ which for typical parameters of long 
GRBs happens between $10^4$ and $10^{5}$ sec.

\section{Slowly rising late-time afterglows of XRFs}

In the CB model, ordinary GRBs are observed mostly
from  viewing angles  $\theta\!\sim\! 1/\gamma$
relative to the CBs' direction of motion.
X-ray flashes (XRFs) are ordinary GRBs  
observed from much larger viewing angles, 
$\theta > 3/ \gamma$ and then $\delta < \gamma/5$ 
(Dar \& De Rujula 2000; DDD2004). 
Such small Doppler factors of XRFs
compared to $\delta\!\sim\!\gamma$ in ordinary GRBs, 
yield $E_{iso}\propto \delta^3 $ smaller by roughly two orders of 
magnitude, $(1+z)\,E_p\!\propto \delta$ smaller by roughly a factor 
$< 1/5$,  and lightcurves in the observer frame that are stretched in 
time by  a factor  $\gamma/\delta > 5$, compared to their respective 
values in ordinary  GRBs.
The unabsorbed
spectral energy density of their emitted SR is given by (DDD2009a and
references therein),
\begin{equation}
F_\nu[t] \propto
           \gamma(t)^{3\,\beta-1}\, \delta(t)^{3+\beta}\,
\nu^{-\beta}\,,
\label{Fnu}
\end{equation}
where $\beta_X\!\approx\!1.1 $ and $\beta_{opt}(t=0)\!\sim \! 0.5$. 
Relativistic energy-momentum conservation yields
the deceleration-law of  CBs 
of a baryon number $N_{B}$ and a radius $R$ in 
an ISM with a constant density $n$ (DDD2009a and references therein):
\begin{equation}
\gamma(t) = {\gamma_0\over [\sqrt{(1+\theta^2\,\gamma_0^2)^2 +t/t_0}
          - \theta^2\,\gamma_0^2]^{1/2}}\,,
\label{goft}
\end{equation}
with $t_0={(1\!+\!z)\, N_{_{\rm B}}/ 8\,c\, n\,\pi\, R^2\,\gamma_0^3}\,.$
Thus, the shape of the entire lightcurve of the SR afterglow,
after entering the constant-density ISM, depends 
only on three parameters, the  product $\gamma_0\, \theta$,
the deceleration time scale $t_0$ and the spectral index $\beta$.

Since $\gamma(t)$ decreases with time, it follows from
Eq.~(\ref{Fnu}) has a maximum when 
$\gamma(t)\,\theta\!=\! [(4\,\beta \!+\!2)/(4\,-\!2\,\beta)]^{1/2}$.  
If the CB enters the constant-density ISM with a larger initial value 
of $\gamma(t)\,\theta$, 
then $F_\nu[t]$ increases slowly as function of time
until  $\gamma(t)\theta$ has decreased to this value.
In the X-ray band $\beta_X\!\approx\! 1.1$ and the peak value of  
$F_\nu[t]$ is reached when $\gamma(t)\theta$ decreased to 1.89 .
In the optical band, initially $\beta_{opt}$=0.5 and the peak value of
$F_\nu[t]$ is reached when $\gamma(t)\theta$ decreased to 1.21 .
Thus after their prompt emission phase, XRFs whose
$\gamma(0)\,\theta > 3$ exhibit a slowly 
rising AG if the progenitor's wind/ejecta has not slowed down the
CBs such that $\gamma(t)\, \theta$ crossed below the above peak values
upon entering the constant density ISM. 

Note that slowly rising X-ray and optical afterglows are not limited to 
XRFs. They can be found also in 
relatively `soft' GRBs with  $1.89\!<\!\gamma(0)\,\theta\!<\!3$. 
GRBs 091029 and 091127 whose lightcurves are shown in 
Figs.~\ref{f6},\ref{f7},\ref{f8} may be such cases.

As can be seen from Eq.~(\ref{goft}), $\gamma$  and $\delta$ in a constant 
density ISM change little as long as $t\!\ll\! t_b\!=\![1\!+\gamma_0^2\,
\theta^2]^2\,t_0$ and Eq.~(\ref{Fnu}) yields the {\it `plateau'}
or shallow decay phase of canonical AGs.
For $t\!\gg\!t_b$,  $\gamma$ and $\delta$ decrease like $t^{-1/4}$
and the AG has an asymptotic power-law decay,
\begin{equation}
F_\nu[t] \propto t^{-\beta-1/2}\,\nu^{-\beta}\, .
\label{Asymptotic}
\end{equation}
The transition $ \gamma_0\! \rightarrow\!
\gamma_0\,(t/t_0)^{-1/4}$
around $t_b$
induces a bend in the AG, the so called `jet  break'.
The post break power-law decline of the AG
(Eq. \ref{Asymptotic}) is
independent of the values of $\gamma_0\, \theta$ and $t_0$.
Because of the relatively small value of $\gamma$ 
at late time, the X-ray and optical bands are both 
well above the bend 
frequency and then $\beta_O\!\approx \! \beta_X\!\approx\! 1.1$,
yielding the same asymptotic decline, 
$F_\nu[t]\!\sim\! t^{\!-\!1.6}\,\nu^{\!-\!1.1},$
in both the X-ray and optical bands.

\section{Comparison with observations}

In Fig.~\ref{f0} we compare the measured equivalent isotropic energy 
$E_{iso}\!=\!(1.1\!\pm\! 0.1)\times 10^{53}$ erg and the rest frame peak 
energy $(1\!+\!z)\, E_p\!=\!222^{+81}_{-36}$ keV of GRB081028, and those 
of other ordinary GRBs and XRFs with known redshift. The $[E_p, E_{iso}]$ 
correlations predicted by the CB model (DDD2007 and references therein) 
for LGRBs and SHBs are indicated by the thick lines. As can be seen, the 
measured values of $E_{iso}$ and $E_p$ in GRB081028 are normal for 
ordinary GRBs and satisfy the predicted CB model correlation, while XRFs 
which are GRBs with small $E_p$ have much smaller values of $E_{iso}$ 
than those of ordinary GRBs. We have also indicated in Fig.~\ref{f0} their 
values for a recently measured `soft' GRB 091127.

In Fig.~\ref{f1} we compare the Swift BAT 15-150 keV lightcurve of 
GRB081028 (Margutti et al.~2009) and its CB model description in 
terms of 2 prompt ICS pulses, each one described by Eq.~(\ref{ICSlc}) with 
$E_p(t)$ as given by Eq.~{\ref{PeakE}) with $E_p(t_p)$=$E_p$ and the $E_p$ 
values reported by Margutti et al.~2009.
In order to minimize the number of adjustable parameters in the CB model
description we have assumed $\beta_g\!=\!0$, which results in a
simple cut-off power-law 
behaviour of the spectrum of the ICS
pulses/flares with a cutoff energy equal to $E_p$ (see Eq.~(\ref{ICSlc})).
The normalization constant $A$, 
the beginning time $T_i$ and the width $\Delta(E)$ at the BAT mid-band 
energy $E$=82.5 keV, and the peak energy $E_p$=$E_p(t_p)$ of each pulse, 
which were used in the CB model description of the lightcurve, are listed 
in Table~\ref{t1}.

In Fig.~\ref{f2} we compare the  0.3-10 keV lightcurve of 
GRB081028 reported in the Swift/XRT light curve repository
($http://www.swift.ac.uk/xrt_curves/$, Evans et 
al.~2009) and its CB model description as a sum of 3 
early-time ICS flares (ICFs) as given by Eq.~(\ref{ICSlc}) and a late-time 
synchrotron radiation flare (SRF) as given by Eq.~(\ref{SRF}). In order to 
minimize the number of parameters, we assumed that after their prompt 
emission, the deceleration of the two leading CBs and their expansion 
merged them into a single leading plasmoid (CB).
The best fit parameters used in
the CB model description are listed in Table~\ref{t1}.
The CB description of the lightcurve during the orbital gap in the Swift
XRT data between 850 and 4130 sec is highly uncertain
and we have plotted only the rise and tail of the third ICF.  
The entire CB model lightcurve has $\chi^2/dof\!=\!1.31$ for $dof$=499.
Note in particular that the  power-law index
of the late-time decay of the SR flare satisfies well the CB model
relation $\alpha_X\!=\!\beta_X\!+\!1=\!\Gamma_X$
with $\Gamma_X=$ 2.091 (+0.063, -0.060) that was reported
in the Swift XRT lightcurve repository (Evans et al.~2009).

In Fig.~\ref{f3} we compare the evolution of $E_p$
during the fast decay of the prompt emission 
in GRB081028 and its CB model description
as given by Eq.~(\ref{PeakE}) with $k$=3.
The  value of $k$ in individual pulses may actually be $\sim$2. 
The value $k$=3 is probably an effective value for the sum 
of ICS flares whose $E_p$ values decrease rapidly with time
as the activity of the central engine weakens and produces smaller 
Lorentz factors.

In Fig.~\ref{f4} we compare the effective photon spectral index that was 
inferred by Margutti et al.~(2009) from the XRT data on GRB081028
in the 0.3-10 keV 
energy range and that inferred from the CB model description, 
Eq.~(\ref{GammaICF}), at mid energy $E$=5.15 keV. Due to 
the orbital gap in the Swift XRT 
data, the predicted photon spectral index between 850 and 4130 sec 
is highly uncertain.

In Fig.~\ref{f5} we compare the 0.3-10 keV 
X-ray lightcurve of XRF080707 
which was measured with the Swift/XRT 
and its CB model description in terms of a tail of an ICS flare 
and a rising late-time afterglow of a GRB viewed far off axis. 
The parameters of the CB model description are listed in table 2.
  
In Fig.~\ref{f6} we compare the 0.3-10 keV X-ray lightcurve of the soft 
GRB091029 which was measured with the Swift/XRT and its CB model 
description in terms of a tail of two ICS flares, a rising late-time 
afterglow of a GRB viewed far off axis and a late time SR flare presumably 
produced during its passage through the boundary of the star formation 
region into the ISM or the halo of the host galaxy. The parameters of the 
CB model description are listed in table 2.

In Fig.~\ref{f7} we compare the 0.3-10 keV X-ray lightcurve of the soft 
GRB091127 ($(1+z)\, E_p\!=\!54\!\pm\!3$ keV; Wilson et al. GCN 10204) that 
was measured with the Swift XRT (XRT light curve repository, Evans et al. 
2009) and its CB model description in terms of a slowly rising afterglow 
of a `soft' GRB . The parameters of the CB model description are listed in 
table 2. 

In Fig.~\ref{f8} we compare the lightcurve of the R band afterglow of 
the `soft' GRB 091127 as reported in recent GCNs 
(Smith et al. 10192; Updike et al. 10195;
Xu et al. 10196,10205; Klotz et al. 10200,10208; Andreev et al. 10207;
Haislip et al. 10219, 10230, 10249; Kinugasa et al. 10248) and its CB
model description in terms of an early 
flare overtaken by a slowly rising afterglow of a `soft'
GRB with a superimposed light from a supernova akin to SN1998bw 
placed at the burst location. The prompt emission flare is not 
constraint by the single data point. 
The data and the CB model predictions were not corrected for extinction in 
our Galaxy and in the host galaxy.

\section{Summary and conclusions}

In the CB model, most XRF and soft GRBs are ordinary GRBs observed from a 
far off-axis viewing angle, i,e, viewing angles much larger than those 
of typical GRBs. This implies smaller 
value of $(\!1+\!z)\,E_p$, much smaller value of $E_{iso}$, prompt pulses, 
flares and afterglows much more stretched in time, and a slowly rising 
late-time afterglow, which turns into an asymptotic power-law decay 
$\!\sim\! t^{\!-\!\Gamma\!+\!1/2}$, as demonstrated here for XRF080707 and 
the soft GRBs 091029 and 091127.

GRB081028 was observed from space with the BAT and XRT aboard Swift 
and from the ground with optical telescopes. The observations 
provided detailed information on its properties, which
differ from those expected from a far
off-axis GRB. The equivalent isotropic energy $E_{iso}$ and peak 
energy $(1\!+\!z)\,E_p$ of its prompt emission are 
typical of ordinary GRBs and satisfy the [(1+z)Ep,Eiso] correlation
for ordinary GRBs (Fig.~\ref{f0}):  The measured value of $E_{iso}$ 
yields $(1+z)\,E_p\!\sim\! 280$ keV, in good agreement with its 
measured value $(1\!+\!z)\,E_p\!=\!222^{+81}_{-36}$ keV, 
which is much larger than the typical values of $E_p$ 
in XRFs. Its lightcurves are well described by the CB model 
assuming an ordinary GRB:  The BAT 15-150 
keV lightcurve of the prompt emission is well described by 
a sum of two pulses/flares produced by two CBs by ICS of glory 
photons 
(Fig.~\ref{f1}). Its 0.3-10 keV X-ray lightcurve, which begins during the 
second peak, shows a steep decay of the prompt emission accompanied by a 
rapid spectral softening with flares superimposed. It is well described by 
a sum of three X-ray flares produced by ICS of glory photons by the 
electrons in CBs late ejected  by a weakening central engine 
(Fig.~\ref{f2}). The smoothly rising X-ray AG, which follows the steep 
decay, peaks around 22 ks, after which it turns into a power-law decay. 
This behaviour of the AG is well described by a late-time SRF 
(Fig.~\ref{f2}). In particular, the photon spectral index during the late 
time flare remains constant, $\Gamma_X\!=\!2.04\!\pm\!0.06$, as expected 
in the CB model for SRFs. Moreover, the power-law decay of the SRF has  
an index $\alpha$=2.09 which satisfies
within errors the CB model prediction (DDD2009a)
$\alpha_X\!=\!\beta_X\!+\!1=\!\Gamma\!=\!2.091\!\pm\!0.063)$,
Had it been an ordinary synchrotron AG of
a far-off axis GRB, i.e.,  like that of XRFs, its 
temporal 
index would have been $\alpha_X\!=\!\beta_X\!+\!1/2\!=\!\Gamma\!-\!1/2 
\!\approx\!1.6 $ (DDD2009a), independent of viewing angle, in 
contradiction 
with the observed value, $\alpha\!\sim\! 2.1$. These temporal and spectral 
properties of the late-time flare in the AG of GRB081028 and the normal 
values of $E_{iso}$ and $E_p$ of its prompt emission agree well with the 
CB model interpretation of the re-brightening of its afterglow - i.e., a 
late-time SR flare produced by the same jet that produced GRB081028,
presumably during its 
passage through the boundary of the star formation region where 
GRB081028 took place, into the ISM or the halo of 
the host galaxy.

\noindent
{\bf Acknowledgment} The authors are grateful to Raffaella Margutti for 
kindly providing the tabulated data used in Figs.~1,3,4.

{}

\begin{deluxetable}{lllc}
\tablewidth{0pt}
\tablecaption{The parameters of the early-time ICS and late-time
SR flares used in the CB model
descriptions of the Swift/BAT and XRT lightcurves of GRB081028.}
\tablehead{\colhead{flare}  & \colhead{$T_i$ [s]  } & 
\colhead{$\Delta t$ [s]} &\colhead{$E_p$ [keV] }
}
\startdata
BAT ICF1  & 20.26 & 73.12 & 50 \\ 
BAT ICF2  & 144.0 & 78.10 & 65 \\  
XRT ICF1  & 205.5 & 134.6 & 5.14\\ 
XRT ICF2  & 411.2 & 235 & 0.19 \\
XRT ICF3  & 665.2 & --- & 0.14\\
\hline
          & $T_w$ [s] & $t_{exp}$ [s] & $a_w$ [s] \\
\hline
XRT SRF   &  7038  & 21176 & 7056 \\
\enddata
\label{t1}
\end{deluxetable}

\begin{deluxetable}{lllc}
\tablewidth{0pt}
\tablecaption{The parameters of the early-time ICS and late-time
SR afterglow and flares used in the CB model
descriptions of the X-ray lightcurves of XRFs 080707 and  091029.}
\tablehead{\colhead{flare}  & \colhead{$T_i$ [s] } &
\colhead{$\Delta t$ [s]} &\colhead{$E_p$ [keV] }
}
\startdata
XRF080707  ICF1  & 4.18 & 31.3 & 9.0\\
GRB091029  ICF1  & $\sim$ 0 & 38.8 & 9.8 \\
GRB091029  ICF2 & 218.5 & 102.5 & 10.93 \\
\hline
          & $T_w$ [s] & $t_{exp}$ [s] & $a_w$ [s] \\
\hline
GRB091029 SRF & $2.09\times 10^5$ & $ 2.46\times 10^5$  & $9.8\times 
10^4$ \\
\hline
          & $\gamma(0)\, \theta$ &$ t_b$ & $\Gamma_X$ \\
\hline
XRF080707  AG & 3.85 & 2111 & 2.06 \\
GRB091029  AG & 2.35 & 5754 & 2.01 \\ 
GRB091127  AG & 2.42 & 1365 & 2.03 \\
\enddata
\label{t2}
\end{deluxetable}

\newpage
\begin{figure}[]
\centering
\epsfig{file=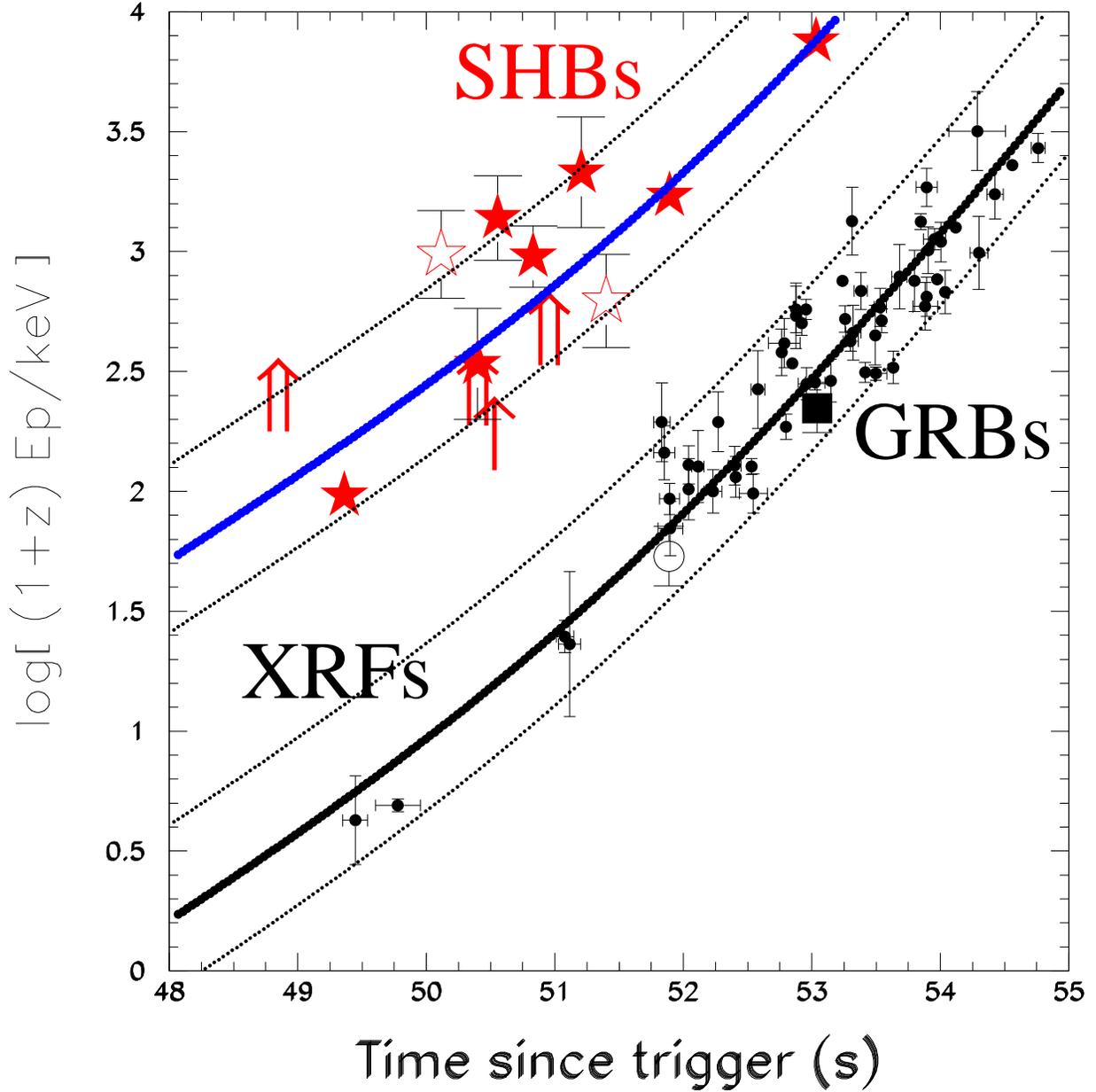,width=18cm}
\caption{
Comparison between the observed correlation $[E_p,E_{iso}]$ in GRBs 
and the correlation predicted by the CB model (thick lines) in LGRBs/XRFs 
(DDD2007, Eq.(4)) and in SHBs
(DDD2009b, Eq.(22)) with known redshift. The long GRB081028 is indicated 
by a full black square.}
\label{f0}
\end{figure}

\newpage
\begin{figure}[]
\centering
\epsfig{file=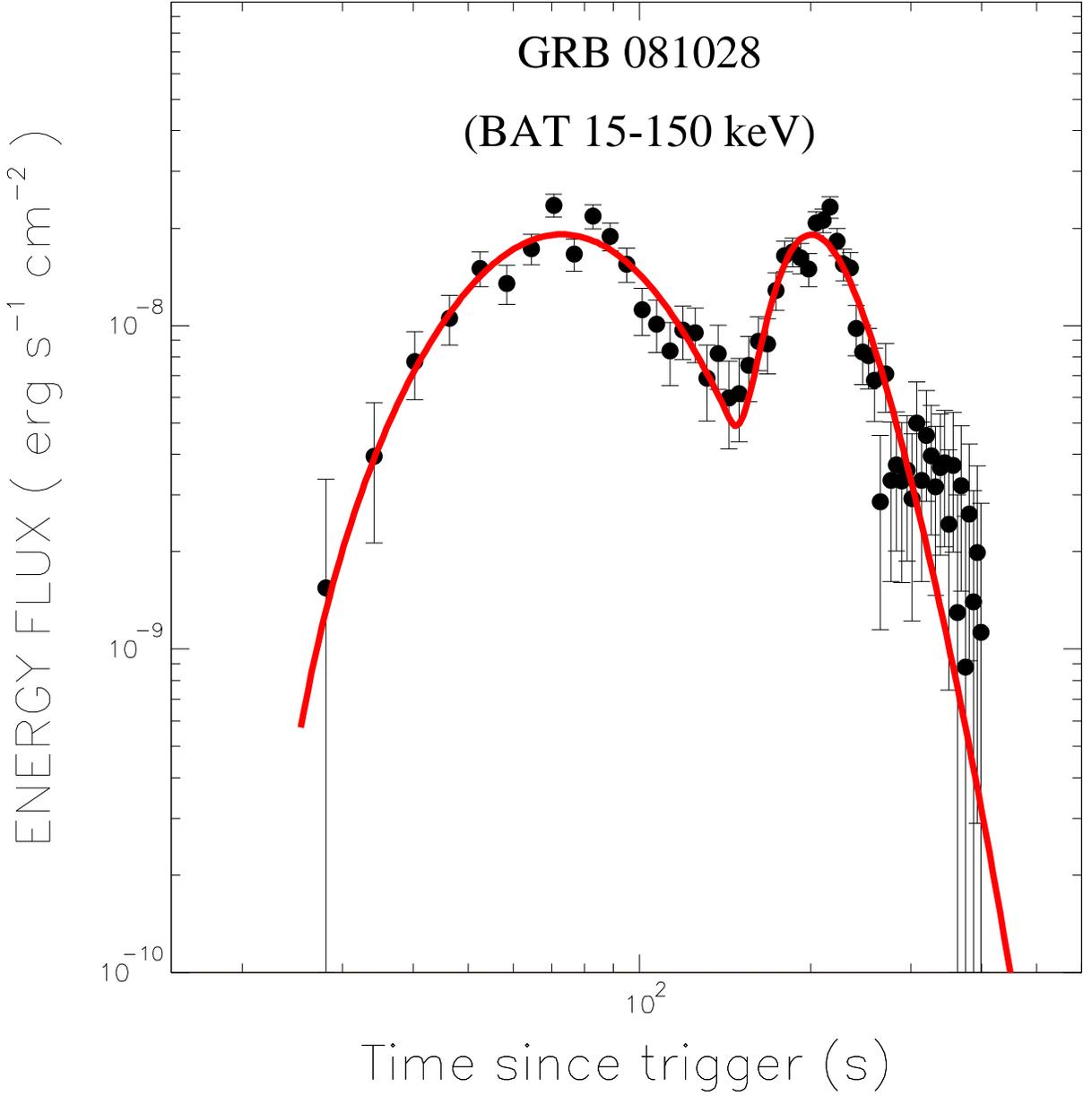,width=18cm}
\caption{Comparison between the Swift BAT lightcurve of
the prompt 15-150 keV emission in
GRB081028 (Margutti et al.~2009)
and its CB model description in terms of two ICS flares
(see the text for details).
}
\label{f1}
\end{figure}
\newpage

\begin{figure}[]
\centering
\epsfig{file=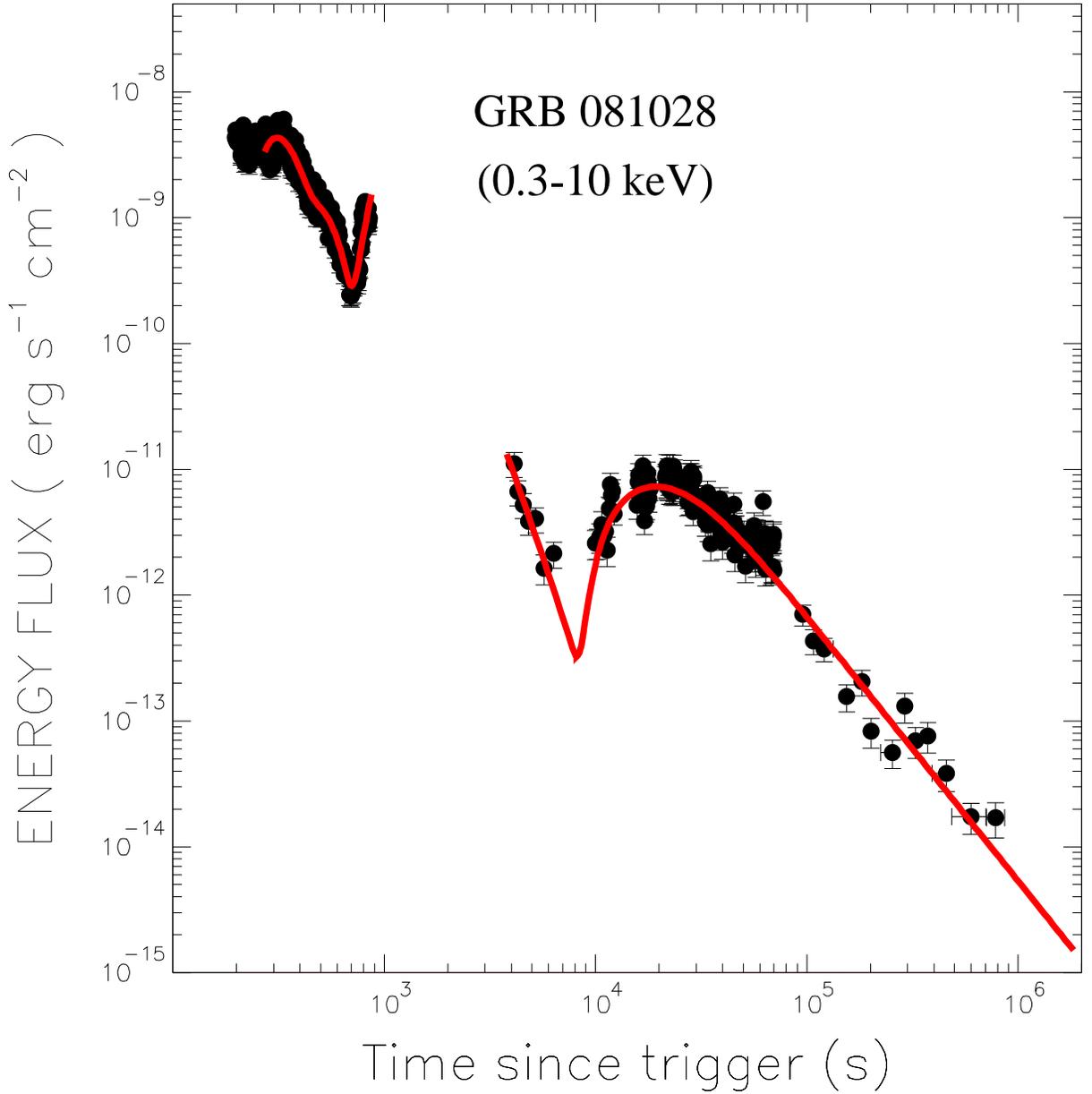,width=18cm}
\caption{Comparison between the Swift XRT  
0.3-10 keV X-ray lightcurve of
GRB081028 (XRT lightcurve repository, Evans et al.~2009)
and its CB model description in terms of 3 early-time ICS flares 
and a late-time SR flare and the parameters listed in table 
(see the text for details).}
\label{f2}
\end{figure}

\newpage
\begin{figure}[]
\centering
\epsfig{file=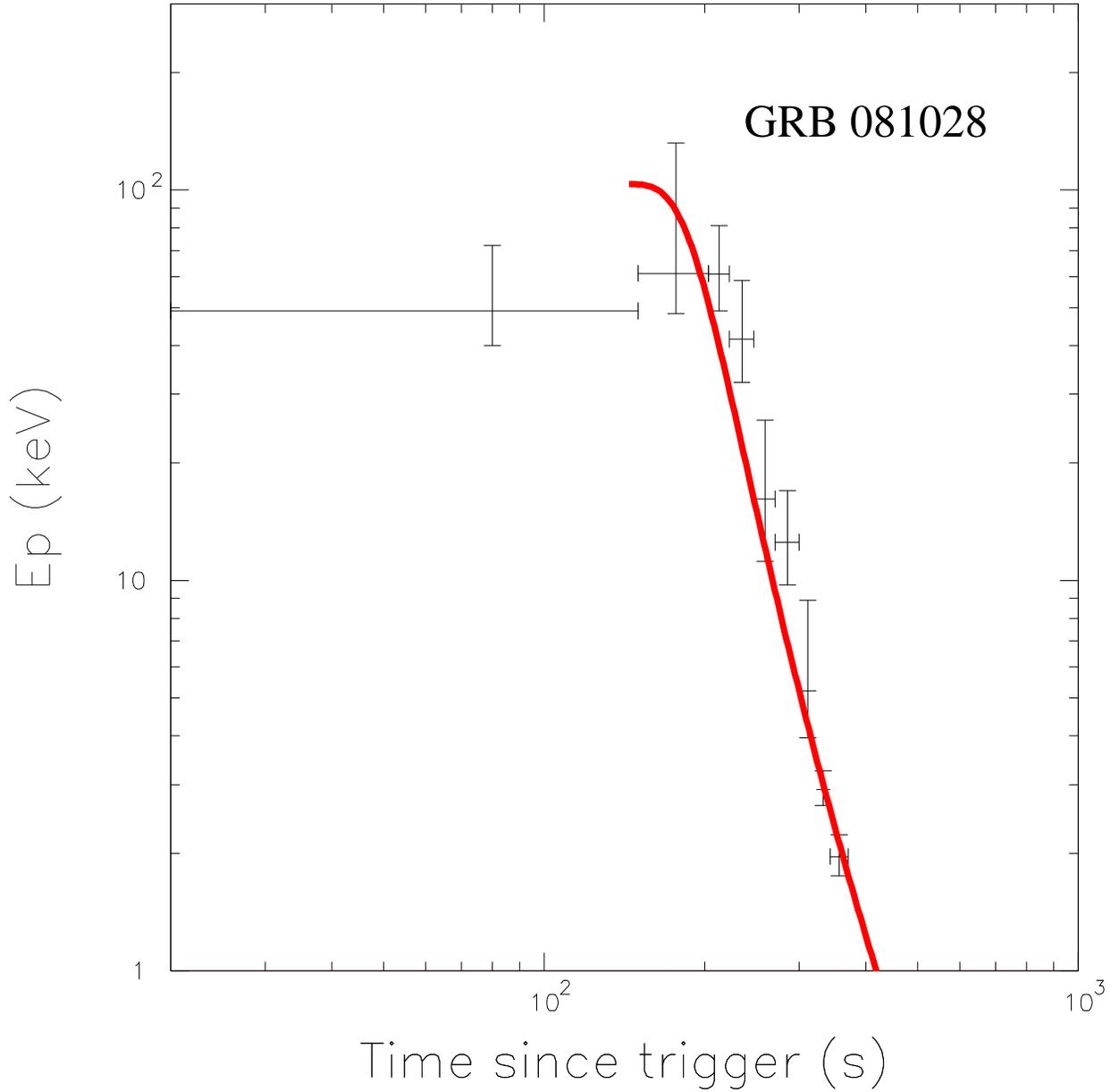,width=18cm}
\caption{Comparison between the temporal variation of 
the peak energy $E_p(t)$ in GRB081028 during the  
fast decay phase of the prompt emission that was inferred 
by Margutti et al.~(2009) from the Swift BAT and XRT observations,
and its CB model description (see text for details).
}
\label{f3}
\end{figure}

\newpage
\begin{figure}[]
\centering
\epsfig{file=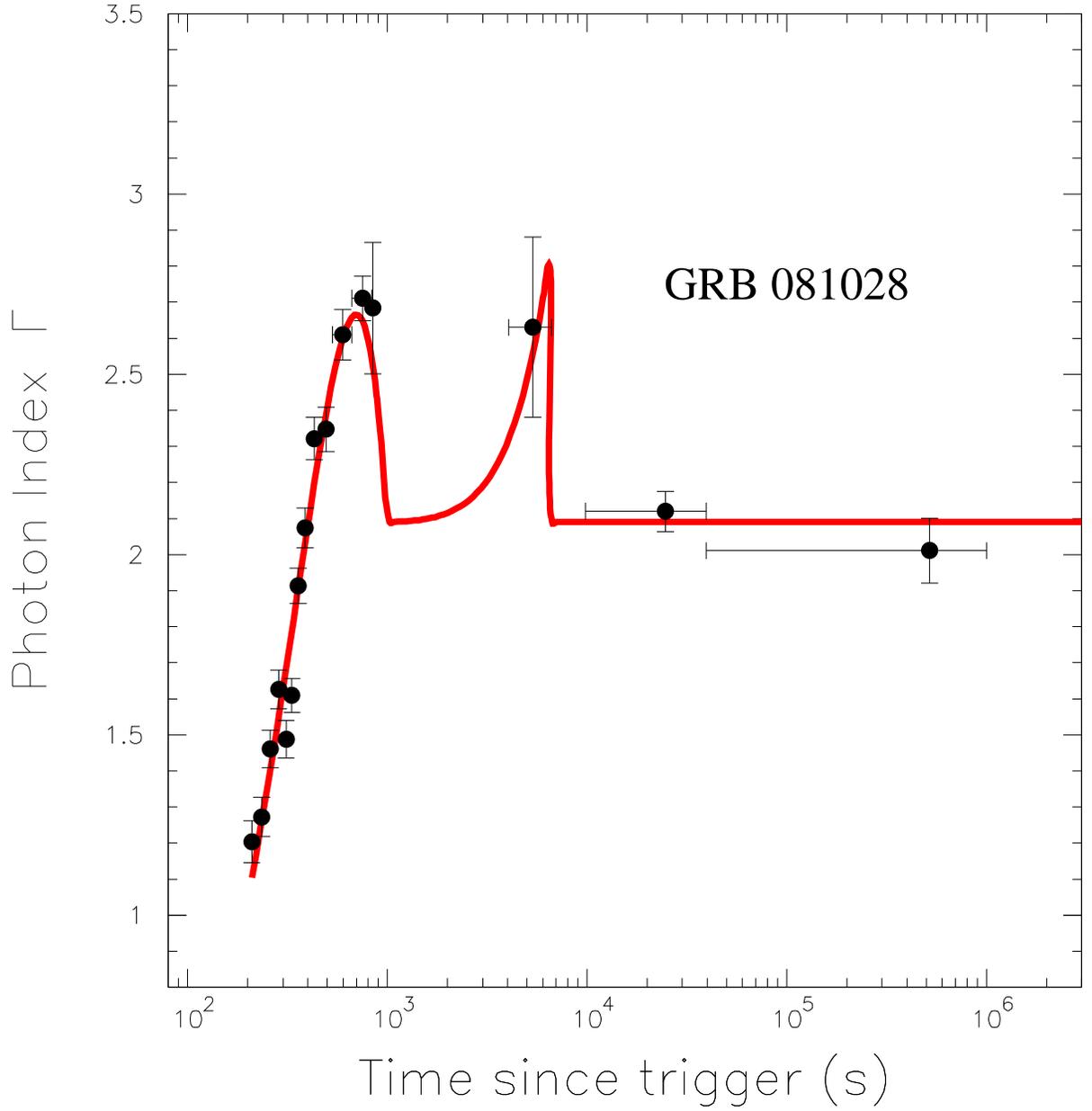,width=18cm}
\caption{Comparison between the effective photon spectral 
index of the 0.3-10 keV lightcurve of GRB081028, which 
was inferred by Margutti et al.~(2009) from the Swift 
XRT observations, and that which follows from the CB model 
description of the XRT lightcurve (see the text for details).
}
\label{f4}
\end{figure}

\newpage
\begin{figure}[]
\centering
\epsfig{file=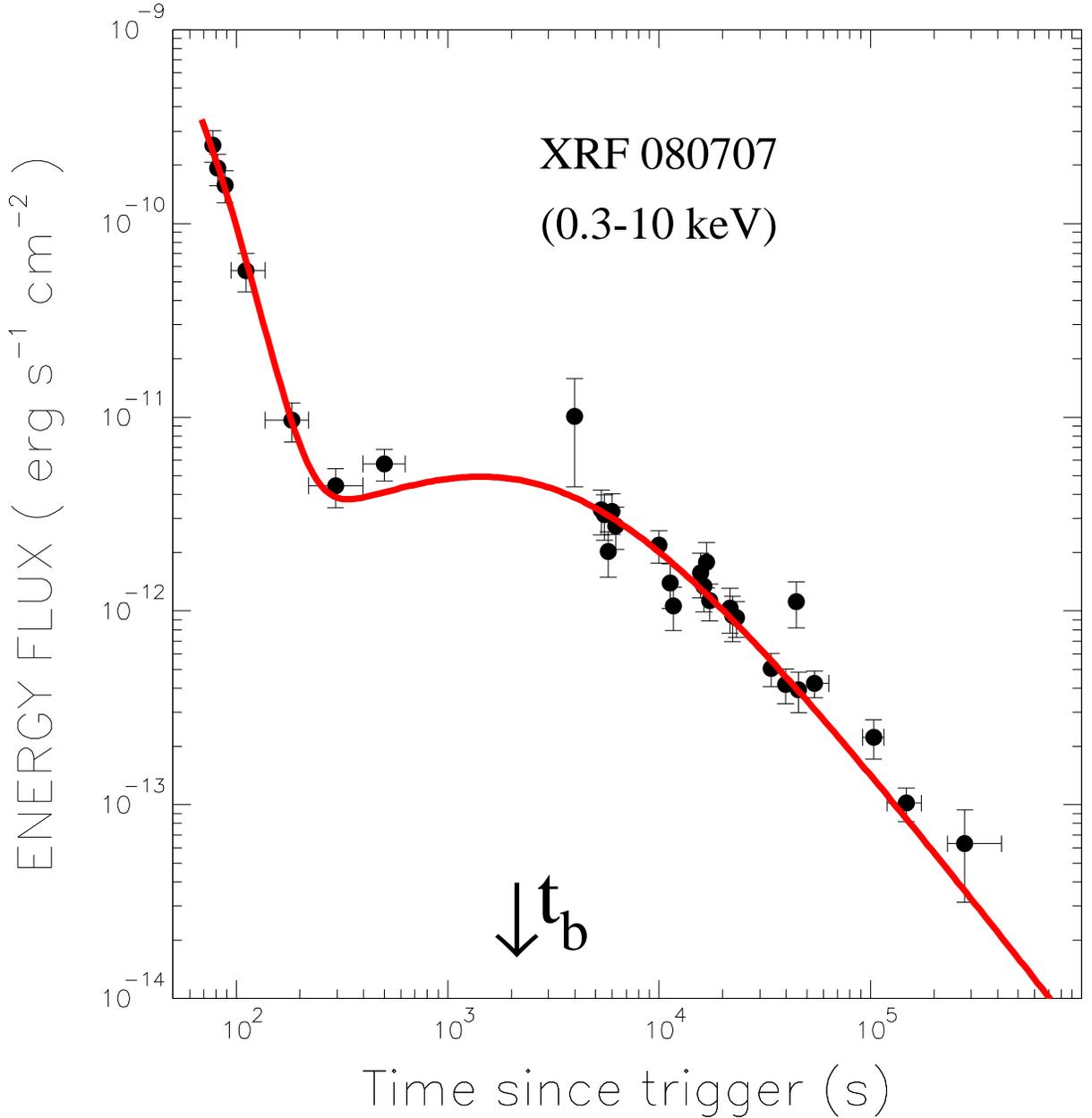,width=18cm}
\caption{Comparison between the 
0.3-10 keV X-ray lightcurve of
GRB081028 reported in the Swift/XRT lightcurve repository 
{\it http://www.swift.ac.uk/xrt$_{-}$curves/},
Evans et al.~2009)
and its CB model description in terms of the tail of an early-time ICS 
flare and a late-time aftergow  assuming a constant density ISM. The 
parameters that were used are listed in Table 2.   
(see the text for details).}
\label{f5}
\end{figure}

\newpage
\begin{figure}[]
\centering
\epsfig{file=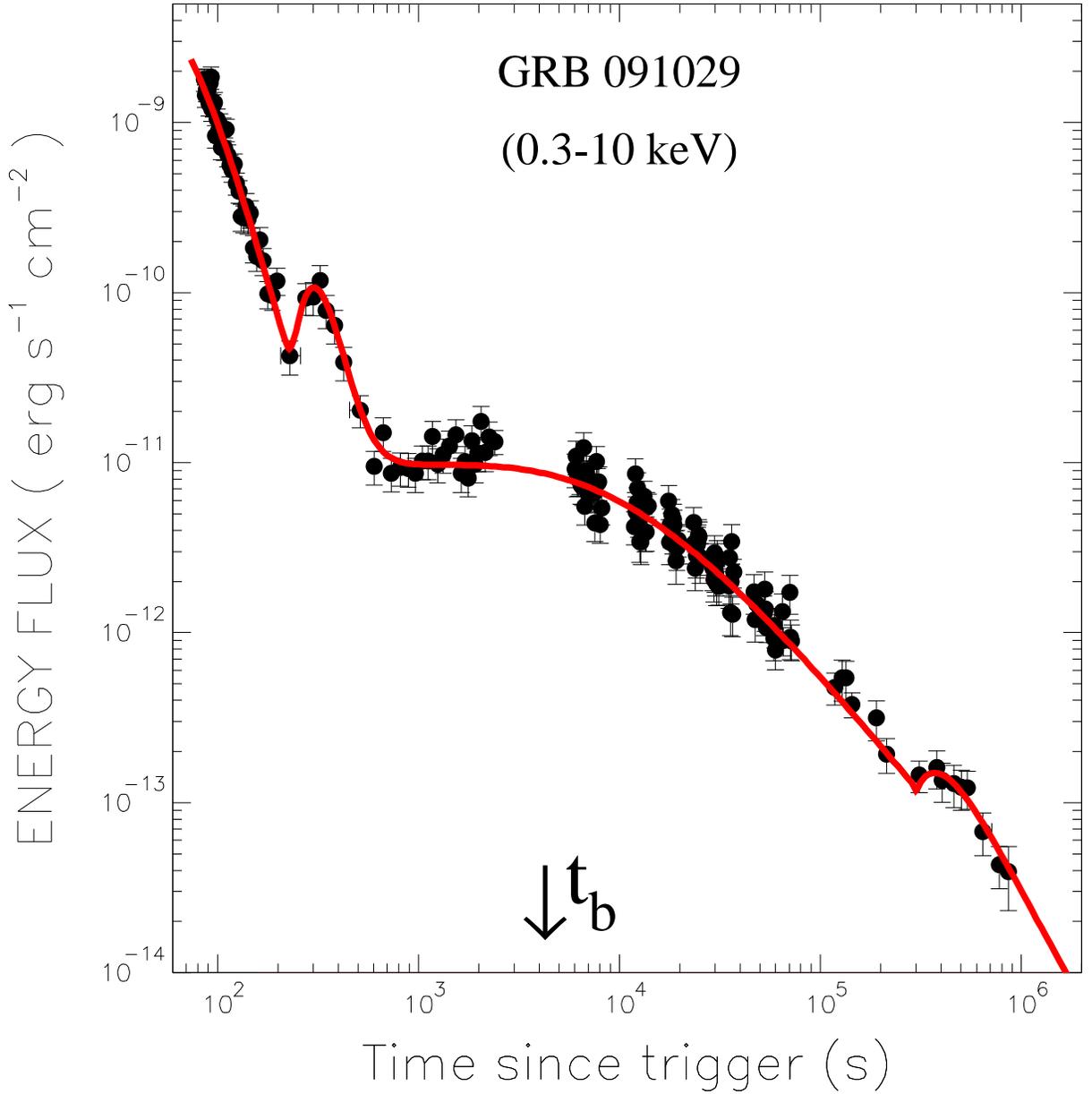,width=18cm}
\caption{Comparison between the Swift XRT
0.3-10 keV X-ray lightcurve of
XRF 091029 (XRT lightcurve repository, Evans et al.~2009)
and its CB model description in terms of two early-time ICS
flarse, a late-time aftergow  assuming a constant density ISM
and a late time SR flare. The
parameters that were used are listed in Table 2.
(see the text for details).}
\label{f6}
\end{figure}

\newpage
\begin{figure}[]
\centering
\epsfig{file=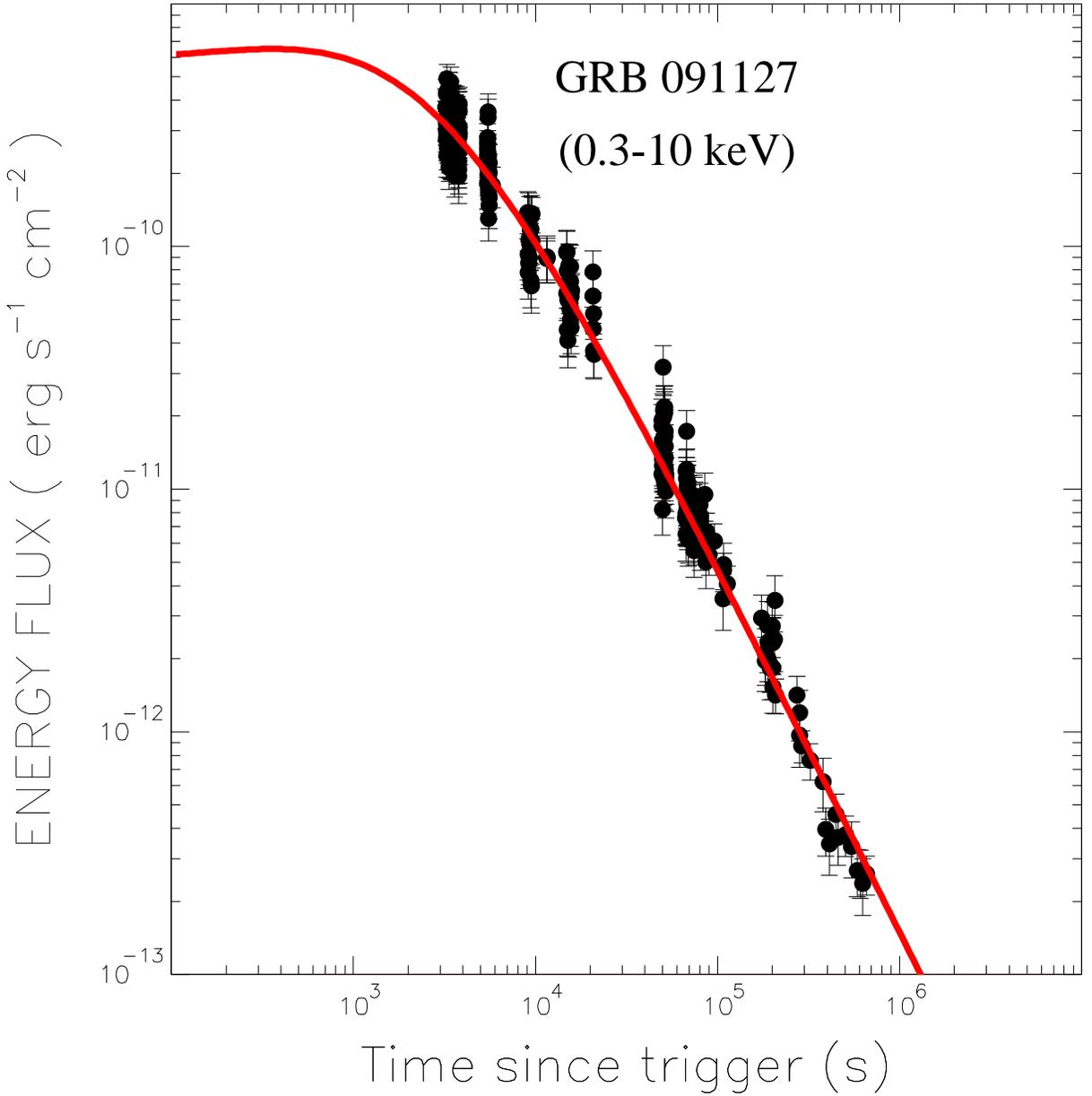,width=18cm}
\caption{
Comparison between the 0.3-10 keV X-ray lightcurve of the soft GRB091127
reported in the Swift/XRT lightcurve repository (Evans et al. 2009) and 
its CB model
description in terms of a rising late-time aftergow. The parameters that 
were used are listed
in Table 2. (see the text for details).
}
\label{f7}
\end{figure}

\newpage
\begin{figure}[]
\centering
\epsfig{file=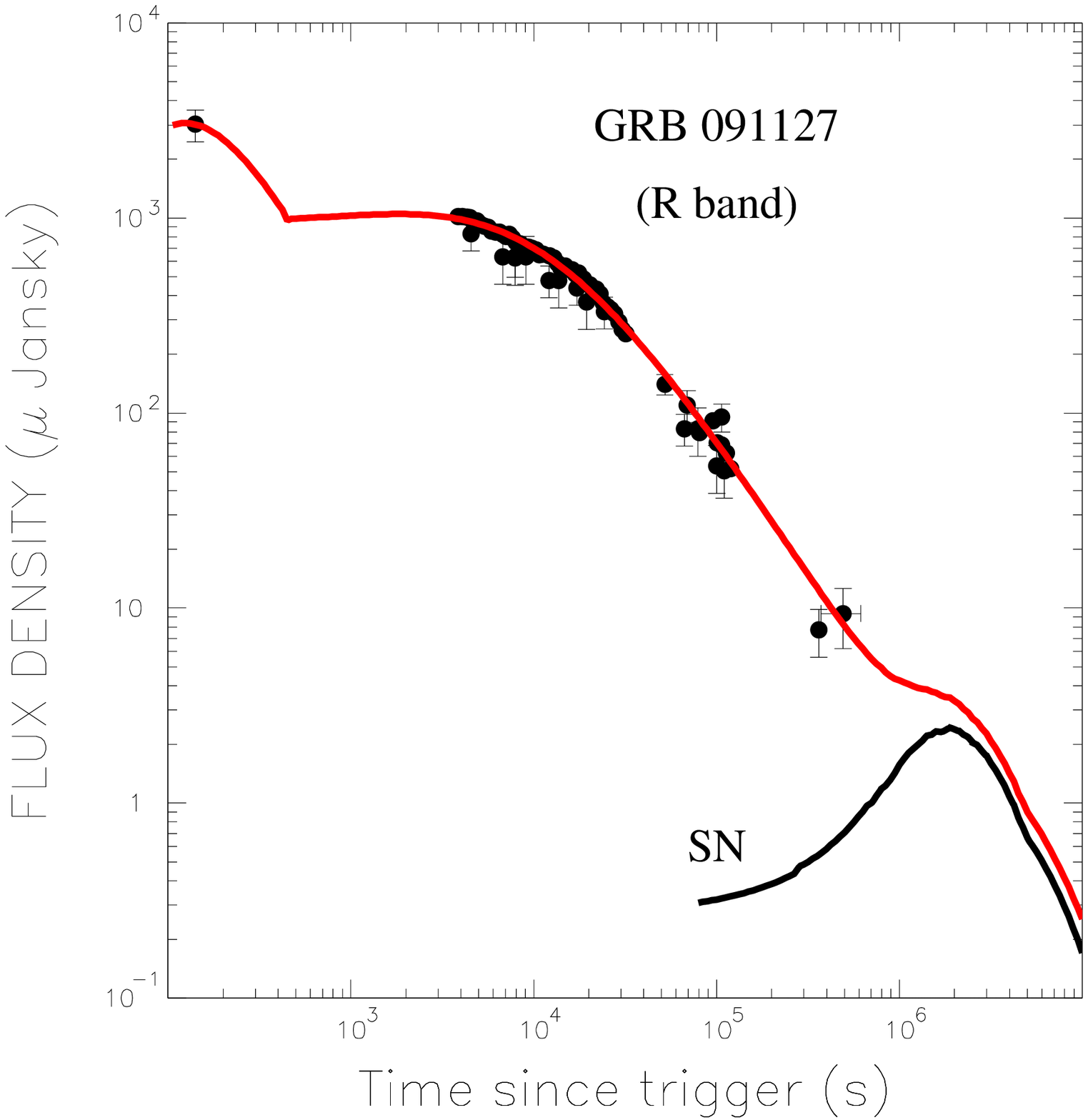,width=18cm}
\caption{
Comparison between the R band light curve of the soft GRB 091127 as 
reported
in recent GCNs (Smith et al. 10192; Updike et al. 10195;
Xu et al. 10196,10205; Klotz et al. 10200,10208; Andreev et al. 10207;
Haislip et al. 10219, 10230, 10249; Kinugasa et al. 10248) and its 
CB model description in terms of a prompt 
optical flare from
the jet collision with the progenitor's wind/ejecta prior to the GRB, a 
slowly rising AG with
a superimposed light from an SN akin to SN1998bw at the GRB location.
The parametes $\gamma(0)\, \theta$, and the late time value of
$\Gamma $ of the rising AG  are those obtained from the CB model
 description of the X-ray lightcurve (see the text for
details).
}
\label{f8}
\end{figure}


\begin{thebibliography}{}


\bibitem{}
Chiang, J. \& Dermer, C. D.~1997, arXiv:9708035

\bibitem{}
Dado, S.,  Dar, A. 2009a,  arXiv:0908.0650

\bibitem{}
Dado, S.,  Dar, A. 2009b,  arXiv:0910.0687

\bibitem{}
Dado, S.,  Dar, A. \& De R\'ujula, A.~2002, A\&A, 388, 1079 (DDD2002)

\bibitem{}
Dado, S.,  Dar, A. \& De R\'ujula, A.~2004, A\&A, 422, 381 (DDD2004)

\bibitem[2007]{DDD2007}
Dado, S., Dar, A. \& De R\'ujula, A.~2007, ApJ, 663, 400 (DDD2007)

\bibitem{}
Dado, S.,  Dar, A. \& De R\'ujula, A.~2008, ApJ, 681, 1408 (DDD2008)

\bibitem[2009a]{DDD2009a}
Dado, S., Dar, A. \& De R\'ujula, A.~2009a, ApJ, 696, 994  (DDD2009a)

\bibitem[2009a]{DDD2009b}
Dado, S., Dar, A. \& De R\'ujula, A.~2009b, ApJ, 693,  311  (DDD2009b)

\bibitem[1998]{Dar1998}
Dar, A.~1998, ApJ, 500, L93

\bibitem[2000]{DD2000}
Dar, A. \& De R\'ujula, A.~2000, arXiv:astro-ph/0008474

\bibitem[2004]{DD2008}
Dar, A. \& De R\'ujula, A.~2004, Phys. Rep. 405, 203


\bibitem{}
Evans, P., et al. 2007, A\&A, 469, 379.

\bibitem{}
Evans et al. 2009, MNRAS, submitted (arXiv:0812.3662)

\bibitem{}
Galama, T. J., et al.~1998, ApJ, 497, L13

\bibitem{}
Mangano, V. et al., 2007 A\&A, 470, 105

\bibitem{}
Margutti, R. et al.~2009,  arXiv:0910.3166

\bibitem{}
Piro, L., et al.~1998, A\&A, 331, L4

\bibitem[1995]{Shaviv1995}
Shaviv, N. \& Dar, A.~1995, ApJ, 447, 863

\end{thebibliography}
\end{document}